\documentclass[article]{jss}

\usepackage{thumbpdf,lmodern}

\usepackage{framed}
\usepackage{comment}
\usepackage{amsmath, amsthm}

\author{
	Stefan Böhringer\\
		\href{mailto:s.boehringer@lumc.nl}{S.Boehringer@lumc.nl}\\
   \And Jesse J. Swen\\
		\href{mailto:j.j.swen@lumc.nl}{J.J.Swen@lumc.nl}\\
}\Plainauthor{Stefan Böhringer, Jesse J. Swen}

\title{Plausibility: Exact inference in \proglang{R}}
\Plaintitle{Plausiblility: exact inference in R}
\Shorttitle{Plausibility in \proglang{R}}

\Abstract{
  Plausbility is a theoretical framework that allows to conduct exact inference in general parametric families. We introduce R-packages \pkg{plausibility} that implements this framework for a wide class of regression models. Plausibility can also be used to test penalized regression models such as estimated by package \pkg{glmnet}. We illustrate the package using a number of R data sets Through a class-based mechanism, the package can be easily extended. We illustrate and discuss computation aspects of the implementation and their impact on real-data analysis.
}

\Keywords{JSS, style guide, comma-separated, not capitalized, \proglang{R}}
\Plainkeywords{JSS, style guide, comma-separated, not capitalized, R}

\Address{
  Stefan Böhringer\\
  Biomedical Data Sciences
  \emph{and}\\
  Clinical Pharmacy\\
  Leiden Univsersity Medical Center\\
  Leiden, The Netherlands\\
  E-mail: \email{s.boehringer@lumc.nl}\\
  URL: \url{http://s-boehringer.org/}\\
  \\
  Jesse J. Swen\\
  Clinical Pharmacy\\
  Leiden Univsersity Medical Center\\
  Leiden, The Netherlands\\
  E-mail: \email{J.J.Swen@lumc.nl}
}

\begin{document}

%% -- Introduction -------------------------------------------------------------

%% - In principle "as usual".
%% - But should typically have some discussion of both _software_ and _methods_.
%% - Use \proglang{}, \pkg{}, and \code{} markup throughout the manuscript.
%% - If such markup is in (sub)section titles, a plain text version has to be
%%   added as well.
%% - All software mentioned should be properly \cite-d.
%% - All abbreviations should be introduced.
%% - Unless the expansions of abbreviations are proper names (like "Journal
%%   of Statistical Software" above) they should be in sentence case (like
%%   "generalized linear models" below).

\section[Introduction: Count data regression in R]{Introduction: The plausibility framework} \label{sec:intro}

In contrast to asymptotic statistics, exact statistical procedures exhibit desirable statistical properties such as type I error control or coverage probabilities for any finite sample size. Exact inference has a long tradition in the statistical literature. \proglang{R} \citep{r_core_team_r_2021} provides implementations of some well known exact statistical procedures including Fisher's exact test \citep{fisher_interpretation_1922, agresti_survey_1992}, function \code{fisher.test}, and Clopper-Pearson intervals for the binomial distribution \citep{clopper_use_1934, agresti_approximate_1998}, function \code{binom.test} including Barnard's test for one known marginal probability \citep{barnard_new_1945, barnard_significance_1947}, Boschloo's unconditional test \citep{boschloo_raised_1970}, sample size calculations \citep{suissa_exact_1985} and paired data tests \citep{berger_exact_2003}. Package \pkg{Exact} contains a comprehensive implementation around Fisher's test (\cite{calhoun_exact_2022}). \pkg{exact2x2} offers exact confidence sets \citep{fay_combining_2015, gabriel_boundary-optimized_2018, fay_confidence_2021, fay_practical_2021} including a modification introduced by Blaker \citep{blaker_confidence_2000}. Function \code{chisq.test} can be used to approximate exact inference for contingency tables with many degrees of freedom using the \code{simulate.p.values = TRUE} option. However, an implementation for a wider class of models is currently lacking in R.\par

The plausibility framework has been developed to enable exact inference for a wide range of parametric models, including regression models \citep{martin_plausibility_2015, bohringer_exact_2022}.
%Some limitations exist which are discussed later.
The idea of Fisher's test and Clopper-Pearson intervals is to use cumulative probabilities on the probability space to define a rejection or confidence set, respectively. Plausibility applies this principle to a general likelihood including  nuisance parameters and continuous distributions and can also be used in the high-dimensional setting. We provide the package \pkg{plausibility} which implements the plausibility framework for regression models.

This paper is structured as follows. We review the plausibility framework in the second section. In section three, we describe package \pkg{plausibility} and illustrate it by performing analyses on a retinoblastoma (RB) data set. This section also contains recommendation on parameters controlling the computation complexity. Section four describes how the package can be extended for likelihoods not yet covered in the package. We close with a discussion and an outlook.

\section{Plausibility Framework} \label{sec:models}

\subsection{Plausibility} \label{sec:plausibility}

For $\theta \in \Theta$, let $P_\theta$ a parametric family of distributions with likelihood function $L(Y, \theta)$. Likelihood based plausibility (cite) is then based on the following likelihood-ratio (LR):
\begin{equation} \label{eq:T}
	T_{Y, \theta} = L(Y, \hat \theta) / L(Y, \theta).
\end{equation}
Here, $\hat \theta$ is the maximum likelihood estimator (MLE) of $\theta$. The theory can also be developed by setting $L(Y, \hat \theta)$ to one (for details see cite). The plausibility function is defined as:
\begin{equation} \label{eq:pl}
	\mathrm{pl}_Y(A) = \sup_{\theta \in A} F_\theta( T_{Y, \theta} ),
\end{equation}
where $F_\theta$ is the distribution function of $T_{Y, \theta}$ and $A \subset \Theta$. $\theta^* = \theta^*(T_{Y, \theta}) :=\arg\sup _{\theta \in A} F_\theta( T_{Y, \theta} )$ is called the plausibility estimate, when it exists. The plausibility estimate has similar properties to the MLE, notably it converges to the true $\theta$ in probability (cite).

The plausibility function measures the probability mass of all outcomes less or equally likely than the observed data. It is therefore based on cumulative probabilties and can be interpreted as a P-value for observation $Y$ to be sampled from family $P_\theta$. For example, it can be used to construct goodness-of-fit tests against families $P_\theta$.

\subsection{Weighted Plausibility} \label{sec:wplaus}

To allow for model comparisons, the plausibility function can be modified to include a weighting function $w$ which is assumed to be independent of $\theta$.
\begin{equation} \label{eq:plw}
	\mathrm{pl}^w_Y(A) = \sup_{\theta \in A} F_\theta( w(Y) ).
\end{equation}
$w$ can be seen as an arbitrary test statistic that orders observations $Y$. The cumulative probability according to this statistic is captured by the weighted plausibility function. For model comparisons, an important choice is the likelihood ratio (LR). If the null hypothesis is represented by $\Theta_0 \subset \Theta$ and the alternative by $\Theta_1 \subset \Theta$ with $\Theta_0 \subset \Theta_1$, the weighing function $w(y) = \sup_{\theta \in \Theta_0} l(y, \theta) - \sup_{\theta \in \Theta_1} l(y, \theta) =: l(y, \hat \theta_0) - l(y, \hat \theta_1) $ can be used to compare the two nested models using exact inference. $l$ is the log-likelihood and $\hat \theta$ is the MLE. Note that this function is indeed independent of $\theta$ as the LR can be pre-computed for every possible $Y$, {\it i.e.} $w$ will return the same value for $Y$, irrespective of the $\theta$ used in the plausibility function. This choice of $w$ leads to an exact model comparison that is asymptotically equivalent to the LR-test thus leading to an efficient and exact procedure. Other choices of $w$ are possible and in general any statistic can be used.

In practice, it is not possible to compute the plausibility function exactly either for continuous or discrete distributions. Stochastic integration can be used to approximate the plausibility function. From the stochastic integration perspective, plausibility is closely related to a parametric bootstrap. However, uncertainty in parameter estimates is taken into account which makes a difference in type I error control \citep{bohringer_exact_2022}. In a nutshell, plausibility can be seen as a parametric bootstrap which is supremized over the parameters of the model, {\it i.e.} conceptionally stochatic integration is performed for all parameter values in $A$.

\subsubsection{Importance sampling} \label{sec:importance}

The plausibility functions $\mathrm{pl}_Y(A)$ and $\mathrm{pl}^w_Y(A)$ are supremized over $\theta \in A$. For each fixed $\theta$, stochastic integration has to be performed to evaluate the cumulative probabilities $F_\theta( Y )$ and $F_\theta( w(Y) )$, respectively. This presents a numerical problem, as the sampling variation due to stochastic integration prevents the optimization to succeed as close to the supremum, the variation of the objective function becomes smaller than this sampling variation. To construct a convergent algorithm, the stochastic sample is drawn once at some $\theta_0$. Integrating at some $\theta$ can now be achieved using an importance sampling (IS) correction for {\it e.g.} weighted plausibility:
\begin{equation} \label{eq:is}
	T^w_{Y, \theta}
		\approx
		\frac{1}{M} \sum_j^M
			\frac{ L(Y, \theta)}{ L(Y, \theta_0) }
			I\{ w(Y) > w(Y^{(j)}) \},
\end{equation}
where $Y^{(i)} \sim \mathrm{iid} P_{\theta_0}$ for $M$ stochastic integration samples.

One potential disadvantage of stocahstic integegration is that it is difficult to approximate small P-values which are limited by the number of stochastic samples with statistic $w$ being larger than for the data.

\subsubsection{Sampling from the alternative}\label{sec:sampling}

To approximate small P-values, the same imporance sampling scheme is used, with a twist. As $\theta_0$ can be chosen arbitrarily, it can be set to $\hat \theta_1$, {\it i.e.} the estimate of $\theta$ under the alternative. In this case, the weighting factor $ L(Y, \theta) /  L(Y, \hat \theta_1)$ is small when $\theta$ is different from $\hat \theta_1$ as the data was generated from a parameter value close to $\hat \theta_1$. As a result the weighted plausibility $T^w_{Y, \theta}$ can become much smaller than $1/M$ which is required when P-values can be very small. This is relevant in applications with a lot of mulitple testing such as genetic studies.\par

In practice, it is desirable to not sample from the parameter value $\hat \theta_1$ as standard errors might be large. The sampling position is therefore controlled by a tuning paramter $\gamma$ so that sampling takes place from $\theta_\gamma = (1 - \gamma) \theta_0 + \gamma \hat \theta_1$.

\subsubsection{Alternative weighing functions}\label{sec:weiging}

While choosing the (log-)likelihood ratio as weighing function entails optimality under certain circumstances \citep{bohringer_exact_2022}, $w$ can be any measurable function. This can be used to implement tests for penalized regression models. The weighing function implemented in the package for penalized regresssion is as follows:

\begin{align}
	w^{pen}(\mathbf X, \mathbf Y) =
	\log \left\{
		\prod_i\varphi(Y_i; (\hat \beta_{a0}^T \mathbf X_0  + \hat \beta_a^T \mathbf X_a)_i, \tau)/\prod_i\varphi(Y_i; (\hat \beta_0^T \mathbf X_0)_i, \tau)
	\right\}.
	\label{formula:pen}
\end{align}

Here $\varphi$ is the density function of the outcome distribution, $\beta_{0}$, $\beta_{a0}$ are low-dimensional nuisance parameters etsimatated under the null and alternative, respectively, corresponding to covariates $\mathbf{X}_0$. $\beta_a$ is a high-dimensional parameter vector corresponding to covariates $\mathbf{X}_a$, and $\mathbf{Y}$ is the vector of outcomes. $\tau$ are additional nuisance parameters such as dispersion parameters. $w^{pen}$ is therefore the LR evaluated in the linear predictors as evaluated under the high-dimensional alternative and the low-dimensional null hypothesis. This implies that $w^{pen}$ is motivated on a heuristic basis.

\subsection{Plausibility Regions} \label{sec:regions}

In analogy with confidence intervals, plausibility allows to calculate sets - plausibility regions - with coverage garuantees for the parameter of interest. For parameter vector $\theta$, the plausibility region is definded as:

\begin{eqnarray*}
  \Pi_y(\alpha) = \{ \theta | \mathrm{pl}_y(\theta) > \alpha \}.
\end{eqnarray*}

This region can be interpreted as a confidence set which covers the true parameter $\theta$ with probability $1 - \alpha$. In general, $\Pi_y(\alpha)$ is a disconnected set.\par
Often, an estimation problem involves nuisance parameters, {\it i.e.} the parameter vector is split into two parts $\theta = ( \psi, \lambda )$, where $\psi$ is the parameter (vector) of interest and $\lambda$ contains nuisance parameters. This region covers the parameter (vector) of interest with probaility $1 - \alpha$ when nuisance paramters are estimated simultaneously. We define the the marginal plausibility region as follows:

\begin{eqnarray*}
  \Pi^m_y(\alpha) = \bigcup_\lambda \{ \psi| \mathrm{pl}_y( (\psi, \lambda) ) > \alpha \}
\end{eqnarray*}

Note, that this construction differs from those given previously (cite, cite). The first construction (cite) is not exact, in general, and the second (cite) less efficient than the construction given here. We give a proof of coverage probability in the appendix.

\subsubsection{Computations}

Computationally, plausibility regions are constructed by first evaluating the plausibility function on an equally spaced grid of covariate values chosen to cover a bounding box of the asymptotic confidence interval. Next, function \code{contourLines} is used to compute and select contours of level $\alpha$. If more than two covariates are involved, contours are computed with respect to the first two covariates and indexed by combinations of values for the other covariates. A marginal plausibility region is constructed by unification of contour sets across covariates to be marginalized over. Packages \pkg{sp} and \pkg{maptools} are used for the set operations. Package \pkg{maptools} allows to check for the presence of a point in the plausibility region via the \code{gContains} function. The implications of the current implementation is illustrated in the example section below.

\section{Extending the package} \label{sec:api}

The plausibility package can be easily extended. A general plausibility model is implemented by subclassing either \code{PlausibilityUnweighted} or \code{PlausibilityWeighted} for standard and weighted plausbility models, respectively.
For regression models, a subclass of \code{plausibilityModel} can be created. We show the negative binomial model as an example.

\begin{Code}
setClass('plausibilityModelNegativeBinomial', contains = 'plausibilityModel',
  representation = list(), prototype = list());

setMethod('initialize', 'plausibilityModelNegativeBinomial', function(.Object,
  family = 'negativeBinomial') {
  .Object = callNextMethod(.Object, family);
  return(.Object);
});

glmFitNb = function(this, X, y, offset) {
  r = glm.nb(y ~ . + 0, as.data.frame(cbind(X, y)));
  par = c(r$coefficients, log(r$theta));
  sds = c(sqrt(diag(vcov(r))), r$SE.theta);
  return(list(par = par, sds = sds, model = r));
}

setMethod("plausFitModel", 'plausibilityModelNegativeBinomial', glmFitNb);

setMethod("plausSample", 'plausibilityModelNegativeBinomial',
  function(this, u, lp, parAncil, par) {
    qnbinom(u, exp(last(par)), mu = exp(lp))
})
setMethod("plausDensityS", 'plausibilityModelNegativeBinomial',
  function(this, x, lp, parAncil, par) {
    dnbinom(x, exp(last(par)), mu = exp(lp), log = TRUE)
})
\end{Code}

First, the class is declared. An \code{initialize} methods should be implemented, which, at a minimum, calls the superclass \code{initialize} via \code{callNextMethod}.

The method \code{plausFitModel} fits a model being provided with a design matrix \code{X}, response vector \code{y} and an offset \code{offset}. Normally, a standard regression should be fitted. The formal requirements for this function are to return a list with a parameter estimate (\code{par}), standard errors (\code{sds}) and the model (\code{model}). By convention, the first values of \code{par} correspond to regression coefficients.
Any further components are other parameters that need to be optimized over. In the case of the negative binomial model, the over-dispersion parameter $\vartheta$ is such an additional parameter.
The default implementation fo this method calls R function \code{glm} with the family given at object initialization. AS the negative binomial model cannot be fitted using \code{glm}, here we use \code{glm.nb} from package \pkg{MASS}.
\begin{leftbar}
Note, that the design matrix \code{X} contains a column for the intercept, so that the model formula should be chosen as \code{y ~ . + 0} in most cases.
\end{leftbar}

Two more methods need to be implemented. First, \code{plausSample} needs to produce a random sample, corresponding to paramter vector \code{par}. A pre-sampled vector of uniformly distributed values is provided in argument \code{u}. The unifrom draw is performed in the main plausibility classes and is ment to increase efficiency and reproducibility. The method itself therefore only needs to compute quantiles corresponding to the uniform draws. Further arguments are the linear predictor \code{lp}, the parameter vector \code{par} and additional parameters \code{parAncil}. If the distribution only depends on the linear predictor only this arguments needs to be used to compute the quantiles. In the case of the negative binomial distribution, the overdispersion parameter needs to be take into account which is provided as the last component of \code{par} (accessed through helper function \code{last}). The addtional argument \code{parAncil} holds so-called non-plausible parameters which can be ignored in most cases. We refer to a previous publication for technical background \cite{bohringer_exact_2022}.

The implementation of helper functions is given in appendix \ref{app:nb}.

\section{Examples}

\subsection{Binomial model}\label{ex:binomial}

In this section, we use the R data set \code{mtcars} (Motor Trend Car Road Tests) to illustrate the package. This data set is part of the standard R-distribution and contains data on several characteristics of different car models. All plausibility computations can be performed with function

\begin{Code}
Plausibility(f0, f1 = NULL, data, family, Nsi = 2e3L, ...)
\end{Code}

Standard plausibility is used when setting \code{f1} to \code{NULL}. Otherwise, the interface resembles that of the \code{glm} function. Parameters \code{f0} and \code{f1} are formulas specifying the regression formula, \code{family} specifies the type of regression and \code{Nsi} specifies the number of stochastic integration samples to be used (see section \ref{sec:importance}).

First, variable \code{am} (automatic transmission 0/1) is regressed on \code{mpg} (miles per gallon) using a logistic model.

\begin{CodeInput}
R> data(mtcars)
R> Plausibility(am ~ mpg, data = mtcars, family = 'binomial', Nsi = 1e4L)
\end{CodeInput}

\begin{CodeOutput}
Plausibility: 9.995e-01
Estimate:
`(Intercept)`           mpg
-1.718393e-05  7.907771e-07
\end{CodeOutput}

Observations \code{am} are therefore fully compatible with a binomial model, for parameter estimates close to zero. This result is trivial as any Bernoulli-outcome is best explained by an intercept model alone that specifies the outcome frequency. A more meaningful question is whether variable \code{mpg} explains the outcome signficiantly better than the intercept model alone. This computation is performed as follows when the nested models are replaced for \code{f0} and \code{f1}:

\begin{CodeInput}
R> Plausibility(am ~ 1, am ~ mpg, data = mtcars, family = 'binomial', Nsi = 1e4L)
\end{CodeInput}

\begin{CodeOutput}
Plausibility: 2.751e-03
Estimate:
(Intercept)
 -0.6388407
\end{CodeOutput}

Model parameters are only estimated under the null model. Therefore only an intercept is given. In conclusion \code{mpg} significantly contributes to the predition of \code{am} status. Comparing the plausible model comparison with a standard \code{glm} model we get:

\begin{CodeInput}
R> summary(glm(am ~ mpg, data = mtcars, family = 'binomial'));
\end{CodeInput}

\begin{CodeOutput}
Call:
glm(formula = am ~ mpg, family = "binomial", data = mtcars)

Deviance Residuals:
    Min       1Q   Median       3Q      Max
-1.5701  -0.7531  -0.4245   0.5866   2.0617

Coefficients:
            Estimate Std. Error z value Pr(>|z|)
(Intercept)  -6.6035     2.3514  -2.808  0.00498 **
mpg           0.3070     0.1148   2.673  0.00751 **
---
Signif. codes:  0 ‘***’ 0.001 ‘**’ 0.01 ‘*’ 0.05 ‘.’ 0.1 ‘ ’ 1

(Dispersion parameter for binomial family taken to be 1)

    Null deviance: 43.230  on 31  degrees of freedom
Residual deviance: 29.675  on 30  degrees of freedom
AIC: 33.675

Number of Fisher Scoring iterations: 5
\end{CodeOutput}

The P-value for \code{mpg} is $7.5 \times 10^{-3}$ compared with $2.8 \times 10^{-3}$ for the plausibility computation indicates that the asymptotic approximation is conservative in this analysis. The plausible model comparison does not give optimal insight into parameter estimates and uncertainties about them. To this end, plausibility regions can be used.

\subsection{Plausiblity Regions}

For the model above, the plausibility region can be computed as follows:
\begin{CodeInput}
R> region <- PlausibilityRegion(am ~ mpg, data = mtcars, family = 'binomial', level = .95,
+   Nsi = 1e3L, Napprox = 30L, sigmaScale = 4)
\end{CodeInput}

Parameter \code{level} specifies the level of the plausibility region and defaults to 0.95. The number of stochastic integration samples is specified by \code{Nsi}. The plausibility function is evaluated on a grid of parameter values with \code{Napprox} number of points per dimension. In the example, a two-dimensional grid with 900 points would be evaluated. Finally, \code{sigmaScale} defines the size of the grid in terms of the standard errors of parameter estimates based on a GLM. If the plausibility region cannot be bracketed by the current grid, an error message is produced. In this case \code{sigmaScale} needs to be increased.

The presence of a point in the plausbility function can be checked with functino function \code{isPointInRegion}. Here, we check whether the maximum-likelihood estimate from the \code{glm} model is present in the plausibility region.

\begin{CodeInput}
R> isPointInRegion(c(-6.6035, 0.3070), region)
\end{CodeInput}

\begin{CodeOutput}
[1] TRUE
\end{CodeOutput}

As the region can be disconnected, in principle, package \pkg{plausibility} uses packages from spatial data analysis for region queries. Function \code{gContains} from package \pkg{rgeos} is used for the containment check.

\begin{figure}[t!]\label{fig:region}
\centering
\includegraphics{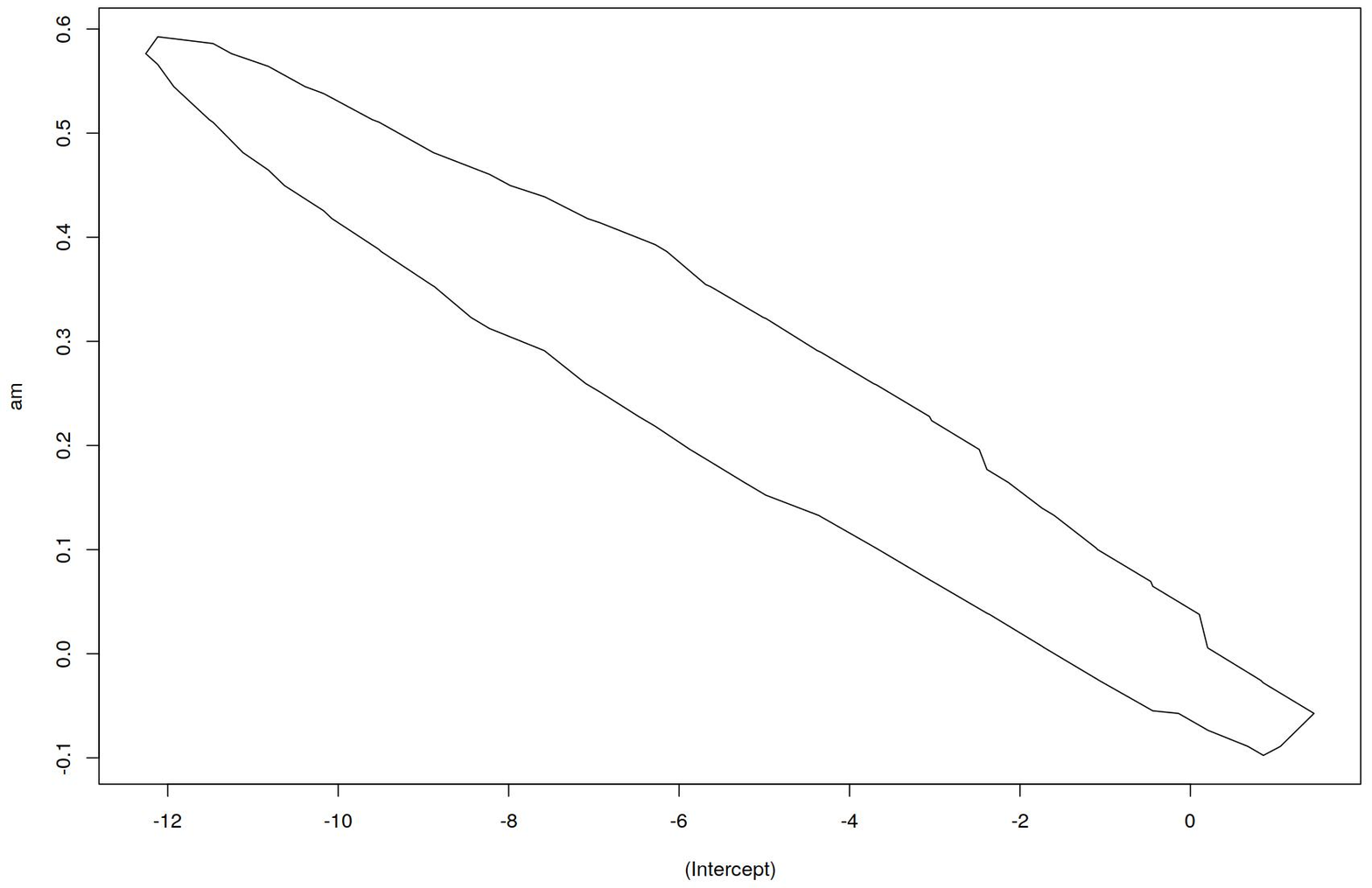}
\caption{Plausibility region for the regression model of the \code{mtcars} data set. {\it Intercept}: x-axis, {\it am}: y-axis.}
\end{figure}

Coordinates of the 95\%-contour of the plausibility function can be extracted as follows.
\begin{CodeInput}
R> coords <- regionCoordinates(region)
\end{CodeInput}

The plausibility region can be plotted with the \code{plot} function and is shown in Figure \ref{fig:region}.

\begin{CodeInput}
R> plot(coords, type = 'l')
\end{CodeInput}

The marginal plausibility region for the regression coefficient of \code{mpg} given the intercept, can be calculated as follows.

\begin{CodeInput}
R> PlausibilityRegion(am ~ 1, am ~ mpg, data = mtcars, family = 'binomial', level = .95,
+   Nsi = 1e3L, Napprox = 30L, sigmaScale = 4)
\end{CodeInput}

The first formula indicates on which parameters to marginalize and the second formula indicates a nested, larger, model to be marginalized. In this case, marginalization results in an interval.

\begin{CodeOutput}
    mpg.lower   mpg.upper 
  -0.09762305  0.59251485 
\end{CodeOutput}

Region plot~\ref{fig:region} indicates that intercept and regression coefficient are negatively correlated. In view of the fact that computation of the plausibility region starts with evaluating plausibilty on a grid of values, the border of the region is not well covered by a grid not well-aligned with the region.

\begin{figure}[t!]\label{sim:regions}
\centering
\includegraphics{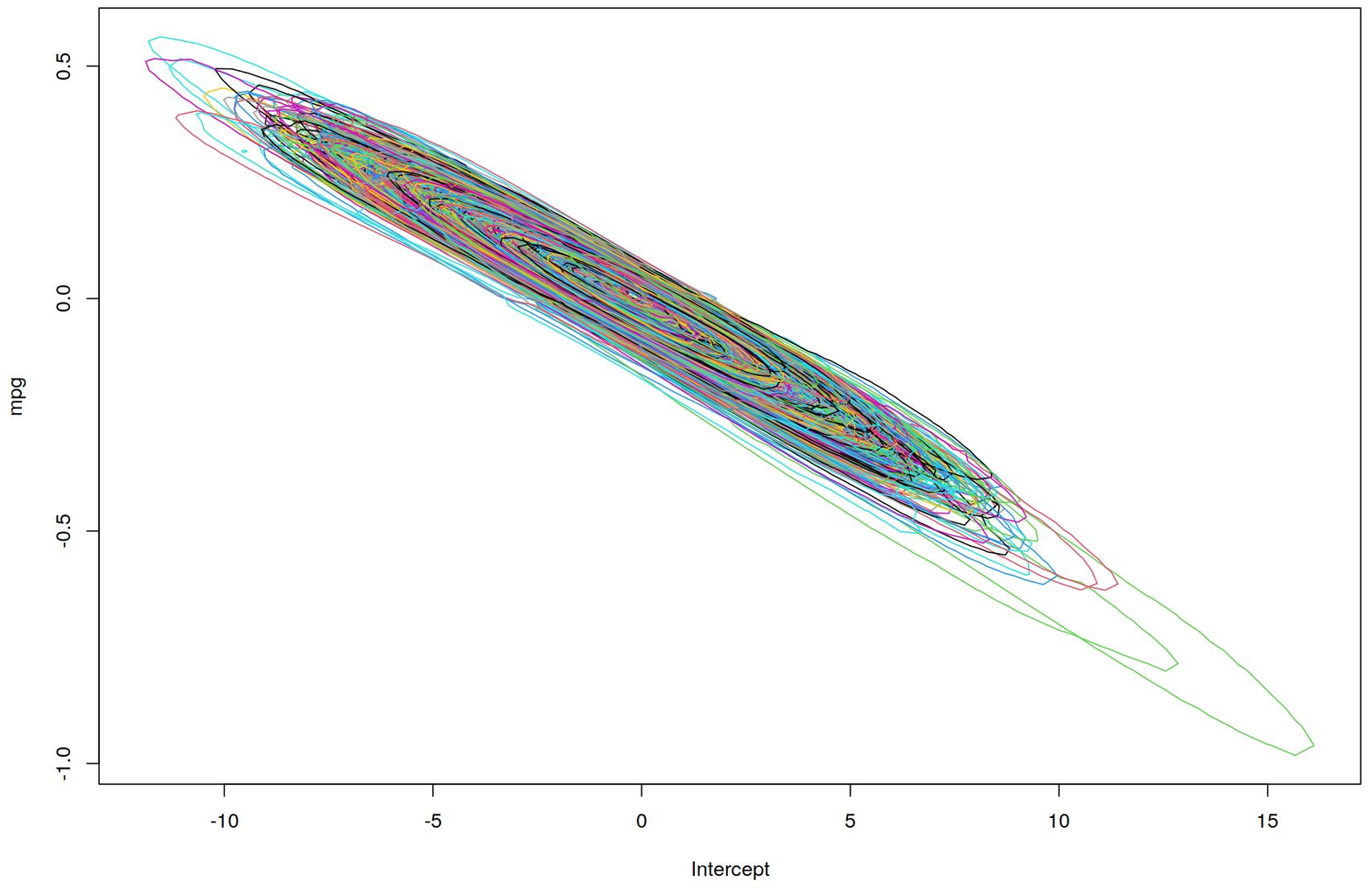}
\caption{\label{fig:region-sim} Plausibility region for the regression model of the \code{mtcars} data set. {\it Intercept}: x-axis, {\it am}: y-axis.}
\end{figure}

Figure~\ref{sim:regions} shows a small simulation of 1,000 replications, where the outcome \code{am} of the \code{mtcars} data set was independently permuted in each replication. Plausibility regions for intercept and \code{mpg} were calculated in each replication.

Coverage of this simulation is 95.1\%, but recalibration was necessary. To account for imprececisions of the grid evaluation, levels of 95\%, 97\% and 99\% are calculated at nominal levels 98\%, 99\%, 99.9\%, respectively. Linear interpolations is used for in-between levels. This behavious is controlled by the \code{calibration} argument to \code{PlausibilityRegion} which defaults to \code{'std'} and implies the behavior above. If set to \code{NULL}, nominal levels will be used directly. We dicsuss improvements to the computation of plausibility regions below.

\subsection{Penalized regression}

Plausibility can also be used to evaluate penalized regression models, where a low-dimensional null-model is compared to a high-dimensional model. To this end, the weighting function $w$ is chosen as the likelihood ratio of the outcome density evaluated under the linear predictors evaluated unter alternative and null, respectively.

\begin{figure}
	\begin{center}
	\includegraphics[width=.9\hsize]{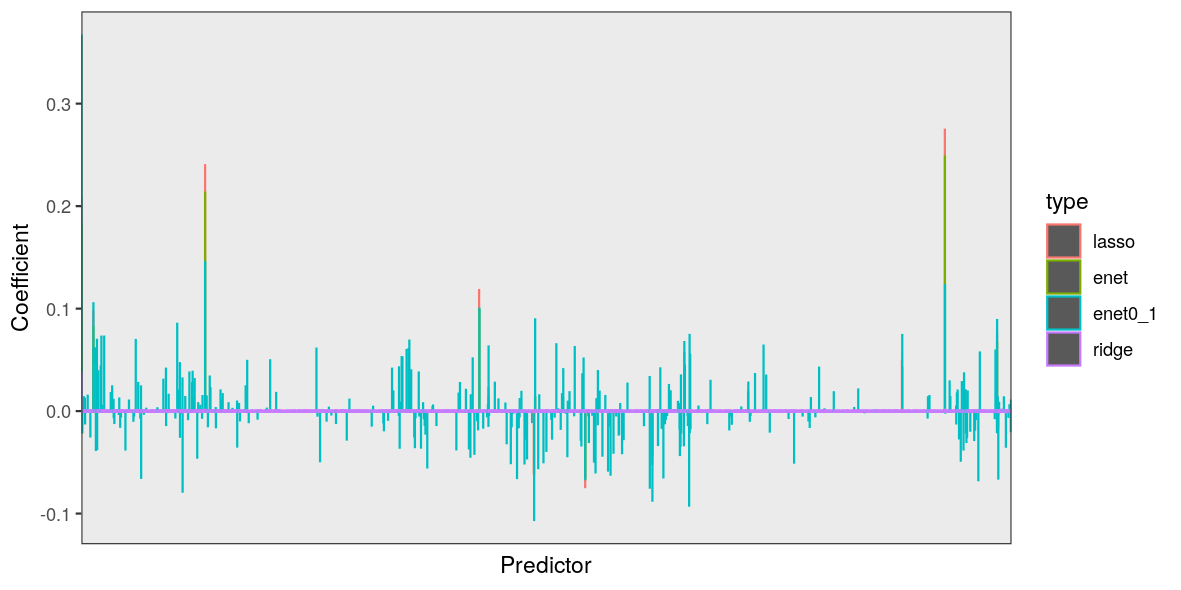}
	\end{center}
	\caption{Penalized regressions. X-axis are predictors 1, ..., 6033 with regression coefficients represented as bars for the methods: Lasso ({\it lasso}), Elastic Net ({\it enet}, $\alpha = 0.9$), Elastic Net ({\it enet0\_1}, $\alpha = 0.1$), Ridge regression ({\it ridge}).}\label{fig:penalized}
\end{figure}

We here analyze a prostate cancer data set %\citea{singh_gene_2002}
as provided by R package \pkg{sda} %\citea{ahdesmaki_sda:_2015}.
The data set contains healthy (N = 50) and prostate cancer samples (N = 52) and measurements of 6033 gene expression values. The analysis is conducted using a logistic model and penalized regression models as implemented by \pkg{glmnet} (\cite{friedman_regularization_2010}). Figure~\ref{fig:penalized} shows regression coefficents as derived from four different models: lasso, elastic net $\alpha = 0.9$), elastic net ($\alpha = 0.1$), ridge, representing models with decreasing sparsity. The penalty parameter was chosen with the \code{cv.glmnet} function using standard settings in each case. On the one extreme, lasso selects just four variables with absolute large values (red bars) whereas ridge selects all variables with absolutely very small coefficients (purple).

We next compute a global p-value for these regression models, where the null-model is the intercept model and the alternative includes all genes.

\begin{CodeInput}
R> dSingh <- get(data(singh2002))
R> dSinghLowDim <- data.frame(y = 2 - as.integer(dSingh$y));
R> pl <- Plausibility(y ~ 1, y ~ ., dSinghLowDim, 'binomial', Nsi = 1e4L,
+   plClass = 'plausibilityPenalized',
+   initArgs = list(X = dSingh$x, NlambdaSel = 50, alpha = 0.5));
\end{CodeInput}

In order to perform the high-dimensional analysis, two data sets have to be provided: (1) the low-dimensional data set used to evaluate the null-model (\code{dSinghLowDim}), (2) the added high-dimensional part (\code{dSingh\$x}). To indicate the high-dimensional nature of the analysis, argument \code{plClass} is set to \code{'plausibilityPenalized'}. Arguments used when constructing a class of this type are specified in argument \code{initArgs}. Here, the high-dimenionsal covariates are passed as argument \code{X} which has to be of class \code{matrix} which is a requirement of function \code{glmnet}. The parameter \code{NlambdaSel} corresponds to a numeric optimization: cross-validation to select penalty parameter $\lambda$ is run this many times and the median is selected. When the stochastic sample is drawn, penalized regression will use this fixed value of $\lambda$ throughout. Other parameters are \code{Nfolds} indicating number of folds of the cross-validation and \code{alpha}, indicating the mixing parameter of function \code{glmnet}.

\begin{table}\centering
    \begin{tabular}{p{2cm}rrr}\label{tab:penalized}
    Name          & $\alpha$    &  P-value (null) & P-value (alt)\\
    \hline
    Ridge           & 0         & $9.99 \times 10^{-5}$  & $2.13 \times 10^{-10}$\\
    Elastic Net     & 0.1       & $9.99 \times 10^{-5}$  &   $10^{-11}$\\
    Elastic Net     & 0.3       & $9.99 \times 10^{-5}$  &   $10^{-11}$\\
    Elastic Net     & 0.5       & $9.99 \times 10^{-5}$  &   $10^{-11}$\\
    Elastic Net     & 0.9       & $9.99 \times 10^{-5}$  &   $10^{-11}$\\
    LASSO           & 1       & $9.99 \times 10^{-5}$  &   $10^{-11}$\\
    \end{tabular}
    \caption{P-values for penalized regression models (see text). Column P-value (null) and P-value (alt) correspond to results where stochastic samples are drawn from the null or alternative distribution, respectively.}
\end{table}

Table~\ref{tab:penalized} shows results for the mixing parameters corresponding to figure \ref{fig:penalized} for two different approaches to P-value approximation. In the first approach (column P-value (null)), stochastic integration samples are drawn under the null hypothesis. In this case, analogously to a bootstrap-procedure, the P-value is approximated as one minus the proportion of times the weighting function exceeds the value of the stochastic sample ( regularized by adding one in nominator and denominator). In this calculation, the size of the P-value is limited by the number of stochastic samples. For column {\it P-value (alt)}, stochastic integration samples are drawn from the alternative hypothesis and re-weighted as described above (section \ref{sec:sampling}).

All plausibility models resulted in P-values $< 10^{-4}$ for the null approach, indicating that all stochastic samples were closer to the null than the observed data. For the alternative approach all P-values were $10^{-11}$ again indicating overwhelming rejection of the null hypothesis. In situations with a high multiple-testing burden, the approximation of small P-values is important and can be achieved with this approach.

%% -- Summary/conclusions/discussion -------------------------------------------

\section{Summary and discussion} \label{sec:summary}

In this paper, we present package {\it plausibility} which allows to perform exact calculations for a wide range of regression models. Our package fills a gap in the R-package landscape as exact calculations have thus far only been implemented for a limited number of statistics. This can be useful for the analysis of small data sets, when asymptotic approximations might be inaccurate. For example, in other work, small pedigree data is analyzed in a genetic application (cite) as an example of data with an underlying population of limited size. The same ideas devloped in these small sample size applications can be applied to the analysis of penalized regression models, which widens the applicability of such models into the evaluation of associations. We believe, that this application is very useful in the analysis, for example, of omics-data.

\subsection{Conceptual aspects}
Although plausibility is presented as allowing exact computations, this notion is limited by the fact that stochastic integration is used. Unlike asymptotic computations, this approximation truely approaches exact values when increasing the number of stochastic integration samples, so that exact values can be approached arbitrarily closely. Still, the number of stochastic samples should be kept in mind as an important parameter. Choosing \code{Nsi} as 500 gives only a first impression in a screening step, but this parameter should be increased to 5,000 or $10^4$ for reliable results. With respect to reproducibility, the packages ensures that sampling is well isolated in the code, so that custom implementations are automatically reproducible by relying on random numbers provided by supporting classes of the package (section \ref{sec:api}). \par

It is also important to note that plausibility is not a ``plug-in'' replacement for standard regression models. Although the concepts translate one-by-one in most cases, there are exceptions. One such example is the plausibility estimate (section \ref{sec:plausibility}). The example on the binomial model (section \ref{ex:binomial}) demonstrates that an intercept only model always has plausiblity of 1 for such data being different, in general, from the data-generating $\theta$. These, so-called non-plausible parameter values are discussed elsewhere (cite) and need to be taken into account when plausibility regions are calculated. In these cases, plausibility regions cannot be used for hypothesis testing. These problems are not present when data is clustered, {\it e.g.} the outcome is binomial within clusters or model comparisons are performed. The genetic example mentioned earlier, uses the number of affected eyes as outcome, where this clustering ensures the existence of the plausibility estimate.

\subsection{Computational ascpects}

Plausibility can be considered a bootstrap ``done right''. Unlike the parametric bootstrap, plausibility accounts for the uncertainty in nuisance parameters estimates by supremizing the statistic over all possible values of these parameters. This entails a high computational burden, as the stochastic integration has to be repeated for many values of nuisance parameters. By using sampling from the alternative (section \ref{sec:sampling}), this burden can be mitigated but not entirely avoided. The stochastic integration step makes the computation of plausibility regions especially challenging. The current implementation starts with a naiv grid search followed by the construction of a level set using R function \code{contourLines} for which the algorithm is not well-documented. Possible improvements involve starting with asymptotic confidence sets and applying a singular-value-decomposition. After rotating the space to align with the singular vectors a more efficient grid-search is possible. Another optimization could be to sub-parition the space after rotation into rectangular regions to avoid searching large areas unlikely to contribute to the borders of the plausibility region. This area remains challenging and we plan to pursue these optimizations in future work.

\subsection{Future work}

At the moment, regression models from the family of generalized linear models have been either implemented or are straightforward to implement. 
 Mixed models form an important class of models and it is interesting to make them available as part of the plausibility package. To this end, algorithmic approaches using EM-algorithms can be used and the integration over latent random effects can be folded into the stochastic integration happening already.
Another aspect concerns non-inferiority analyses. As part of other work, we have investigated non-inferiority tests for binary outcomes that can be implemented as model comparisons (cite) and we plan to integrate this work.
An important limitations concerns the number of nuisance covariates that can be handled in the computations. As a generic optimization algorithm needs to be used due to the non-smooth nature of the plausibility function, such algorithms typically have exponential running time in the number of parameters to optimize over. In practical terms, in the current implementation, at most 10 covariates can be handled. We are looking into potential mitigations, namely using summaries for groups of variables. We are actively working on this topic to broaden applicability of plausibility models.

In conclusion, plausibility \pkg{plausibility} adds several useful statistical methods to the toolbox of the R user. We believe that several relevant areas can profit from these extensions and are actively working on improvements to the package.

%% -- Optional special unnumbered sections -------------------------------------

\section*{Computational details}

% \begin{leftbar}
% If necessary or useful, information about certain computational details
% such as version numbers, operating systems, or compilers could be included
% in an unnumbered section. Also, auxiliary packages (say, for visualizations,
% maps, tables, \dots) that are not cited in the main text can be credited here.
% \end{leftbar}

The results in this paper were obtained using \proglang{R}~4.1.1 with the \pkg{plausibility}~0.6.1
and \pkg{MASS}~7.3.54 packages. Simulation were parallelized using package \pkg{parallelize.dynamic} \citep{bohringer_dynamic_2013}. \proglang{R} itself and all packages used are available from the Comprehensive \proglang{R} Archive Network (CRAN) at \url{https://CRAN.R-project.org/}.

% 
% \section*{Acknowledgments}
% 
% \begin{leftbar}
% All acknowledgments (note the AE spelling) should be collected in this
% unnumbered section before the references. It may contain the usual information
% about funding and feedback from colleagues/reviewers/etc. Furthermore,
% information such as relative contributions of the authors may be added here
% (if any).
% \end{leftbar}

%% -- Bibliography -------------------------------------------------------------
%% - References need to be provided in a .bib BibTeX database.
%% - All references should be made with \cite, \citet, \citep, \citealp etc.
%%   (and never hard-coded). See the FAQ for details.
%% - JSS-specific markup (\proglang, \pkg, \code) should be used in the .bib.
%% - Titles in the .bib should be in title case.
%% - DOIs should be included where available.

%\bibliography{refs}

%% -- Appendix (if any) --------------------------------------------------------
%% - After the bibliography with page break.
%% - With proper section titles and _not_ just "Appendix".

\bibliography{plaus-softw}

\newpage

\newpage

\begin{appendix}

\section{Plausibility regions} \label{app:region}

In this section, we give the construction of marginal plausibility regions and proof exact coverage properties.

We consider paramters $\theta = ( \psi, \lambda )$ where $\lambda$ is considered a nuiscance parameter such that the marginal plausibility region
$\{ \psi | \mathrm{mpl}_y(\psi) > \alpha \}$
is of interest, with
$\mathrm{mpl}_y(A) = \sup_{ \psi \in A } F_{\theta}(T_{y, \psi})$.

Starting with the plaubility region
$ \Pi_y(\alpha) = \{ \theta | \mathrm{pl}_y(\theta) > \alpha \}$,
we define a marginal plausibilty region as
$ \Pi^m_y(\alpha) = \bigcup_\lambda \{ \psi| \mathrm{pl}_y( (\psi, \lambda) ) > \alpha \}$.

\newtheorem{lemma:coverage}[lemmas]{Lemma}
\begin{lemma:coverage}\label{lemma:conv}
	The marginal plausibilty region $ \Pi^m_y(\alpha)$ has nominal coverage probability,\\
	{\it i.e.} $P(\Pi^m_y(\alpha) \ni \psi) \ge 1 - \alpha$.
\end{lemma:coverage}

\begin{proof}
	\begin{align*}
		  &	P(\Pi^m_y(\alpha) \ni \psi)
		=	P(\bigcup_\lambda \{ \psi| \mathrm{pl}_y( (\psi, \lambda) ) > \alpha \} \ni \psi ) \\
		= &	P( \{ (\psi, \lambda) | \mathrm{pl}_y( (\psi, \lambda) ) > \alpha \} \ni (\psi, \lambda) ) \ge 1 - \alpha \\
	\end{align*}
	The last inequality follows by construction.
\end{proof}

\section{Implementation of Negative Binomial Regression} \label{app:nb}

\begin{Code}

glmFitNb = function(this, X, y, offset) {
	r = glm.nb(y ~ . + 0, as.data.frame(cbind(X, y)));
	par = c(r$coefficients, log(r$theta));
	sds = c(sqrt(diag(vcov(r))), r$SE.theta);
	return(list(par = par, sds = sds, model = r));
}
setMethod("plausFitModel", 'plausibilityModelNegativeBinomial', glmFitNb);

s2fromMuSize = function(mu, size)(mu/size + 1)*mu
probFromMuS2 = function(mu, s2)(1 - mu/s2)
probFromMuSize = function(mu, size)probFromMuS2(mu, s2fromMuSize(mu, size))
\end{Code}

\end{appendix}

%% -----------------------------------------------------------------------------

\end{document}